\documentclass[pra,twocolumn,showpacs,preprintnumbers,amsmath,amssymb]{revtex4}

\usepackage{graphicx}
\usepackage{dcolumn}
\usepackage{bm}

\begin{document}

\title{Universal Effective Quantum Number \\
for Centrally Symmetric Problems}

\author{A.~A.~Lobashev}
 \email{lobashev@vniim.ru}
\author{N.~N.~Trunov }
 \email{trunov@vniim.ru}
\affiliation{
D.~I.~Mendeleyev Institute for Metrology\\
Russia, St.~Petersburg, 190005, \\
Moskovsky pr., 19
}

\date{\today}

\begin{abstract}
An effective quantum number determining with high
accuracy the levels ordering in arbitrary centrally symmetric
potentials for any space dimensionality is introduced and
calculated by means of certain universal methods based on the known
estimates for the total number of the bound states in the same potential
for various dimensionality. Coincidence with some known exact results is
demonstrated. The effective number is used for constructing the
periodical system of the atomic electron shells.
\end{abstract}


\maketitle

\section[1]{Introduction}

The lack of analytic solutions for the most centrally symmetric
potentials calls for developing method approximating the spectra.
It becomes extremely actually at present as besides known problems
in atomic physics, theoretical models of nuclei \cite{Bohr} or
quarkonium \cite{Martin86,Quigg,Martin80} there arises a number of new
objects  (metallic clusters \cite{Kotos, Ostr97} etc) for which
some analogies of the periodic system of shells may be constructed
\cite{Ostr2001,Block,FZ}. They may differ not only by the nature of
selfconsistent field but even by their dimensionality $d$.
A lot of endeavours have been put to obtain rigorous results, see e.g.
\cite{Martin86,Quigg,Martin80,Martin87,21,22}, but these
theorems only give some inequalities for special forms of potentials
and only for $d=3$.

A frequent supposition is that the
energy values depend on some linear combinations of the radial $n_r$
and orbital $l$ quantum numbers, i.e.
$E (n_r , l) = E (\alpha n_r + \beta l)$. The Madelung-Kletchkovsky
rule predicts appearing of new shells $(n_r , l)$ in the Periodic
system of the elements
with increasing $n_r + 2 l$ \cite{Ostr2001,Block,FZ}, for the
metallic clusters it is expected $E=E(3 n_r +l) $ \cite{Kotos, Ostr97},
a similar quantum number for nuclei was proposed in \cite{Bohr} by
using some classical analogy.

However, such dependence of the exact spectra on some linear combination
is known only for the Coulomb and oscillator potentials:
\begin{equation}
\label{1}
V_c (r) = - \frac{Z}{r},
\quad
E_c (n_r , l) = - C_c \left( \nu + \lambda \right)^{-2} ,
\end{equation}
\begin{equation}
\label{2}
V_{osc} (r) = b r^2,
\quad
E_{osc} (n_r , l) =  C_{osc} \left( \nu + \lambda /2 \right) ,
\end{equation}
\begin{equation}
\label{3}
\nu = n_r + \frac{1}{2}, \qquad \lambda (d) = l + \frac{d-2}{2} .
\end{equation}

Two remarkable facts are known  (at least for $d=3$) for our reference
potentials (\ref{1}) and (\ref{2}) and only for them.
First of all, the usual WKB condition
\begin{equation}
\label{4}
\frac{1}{\pi} \int \sqrt{2 \left( E - V(r) \right)
- \frac{\lambda^2}{r^2}  }dr = \nu
\end{equation}
leads to the exact spectra (\ref{1}), (\ref{2}). For $d=3$ the term
$\lambda^2 / r^2 $ is known as the centrifugal potential with the
Langer correction $l (l+1) \rightarrow (l+1/2)^2 $ \cite{Langer}.
Besides,  for the oscillator
\begin{eqnarray}
\label{5}
& & \frac{1}{\pi}  \int \sqrt{2 \left( E - V(r) \right)
- \frac{\lambda^2}{r^2}  }dr  = \nonumber \\
&  & =  \frac{1}{\pi}
\int \sqrt{2 \left( E - V(r) \right) }dr  - \frac{\lambda}{2}  .
\end{eqnarray}
In all cases the integration goes between corresponding turning points
and we put $m=\hbar =1$.
Combining (\ref{4}) and (\ref{5}), we obtain
\begin{equation}
\label{6}
\frac{1}{\pi} \int \sqrt{2 \left( E - V(r) \right) }dr =\nu + \frac{\lambda}{2} ,
\end{equation}
so that in the right hand of (\ref{6}) we see the same linear combination
of $\nu $ and $\lambda $ as in (\ref{2}),
a similar situation with $\nu + \lambda $ we obtain for (\ref{1}).

In the present paper we get an effective quantum number
\begin{equation}
\label{7}
T  \left( n_r , l \right)
\equiv \nu + \phi \lambda = \left( n_r  + \frac{1}{2} \right)
+  \phi \left( l  + \frac{d - 2}{2} \right)
\end{equation}
for any centrally symmetric
potentials and any dimensionality of the problem
by means of some generalization and new universal variations of the
methods used earlier \cite{TMP99,TMP00,Basel}
in order to obtain $T$ for $d=3$.
The coefficient $\phi $ is determined as some definite
combination of the functionals $N_d [E ; V]$, which represent asymptotic
estimates for the total number of the bound states in a given potential
$V(r)$ with the energies not exceeding $E$. These estimates are obtained
without using any WKB method.

This quantum number $T$ (\ref{7}) determines the order of the bound
states: $E(T) > E(T')$ if $T > T'$.  For our reference cases (\ref{1})
and (\ref{2}) $T$ coincides with $\nu + \lambda $ and $\nu + \lambda /2$
correspondingly. In a general case $T$ approximates very good the real
situation, but is not formally the exact quantum number.

Note that the principal quantum number $n=n_r + l + 1$
determines the spectrum only for the Coulomb potential if $d=3$.

\section[2]{The Schr\"odinger equation and its conformal transformation}

We use the Schr\"odinger equation in  the form ($m= \hbar  =1$)
\begin{eqnarray}
\label{8}
& & \Delta_d \Phi +P^2  \Phi  = \nonumber \\
& & = \left(
\frac{\partial^2}{\partial r^2}  +
\frac{d-1}{r} \frac{\partial}{\partial r}  \right)
\Phi +
\frac{\Delta_{d-1} (\Omega ) \Phi}{r^2}  + P^2 \Phi =0, \nonumber \\
& & P^2  =  2 \left( E - V(r) \right),
\end{eqnarray}
where $\Delta_{d-1} (\Omega )$ denotes the Laplace operator on the unit
sphere $S^{d-1} $, $E$ is the energy value and $V(r)$ is the potential.
The equation (\ref{8}) corresponds to the following metric:
\begin{equation}
\label{9}
ds^2 = dr^2 + r^2 d \Omega^2_{d-1}
\end{equation}
with $0 \le r < \infty $ and $ \Omega_{d-1} $ being coordinates on
$S^{d-1}$.

In order to eliminate the first derivative in (\ref{8}) and the
singularity at $r=0 $ we represent the Schr\"odinger equation
in a conformal metric $d\tilde{s}^2 $ with a new variable $\rho = \ln r$
so that
\[
ds^2 = r^2 d \tilde{s}^2 = e^{2 \rho } \left( d\rho^2 +
d\Omega^2_{d-1} \right).
\]
Using   [9, \S 28] we get in the new metric instead of (\ref{8}):
\begin{equation}
\label{10}
\frac{d^2 \Psi}{d \rho^2}   + \Delta_{d-1} (\Omega ) \Psi
+ K(d) \Psi + e^{2\rho } P^2 \Psi =0,
\end{equation}
\begin{equation}
\label{11}
K(d) = - (d-2)^2 /4; \qquad
\Psi = e^{\frac{d-2}{2}\rho } \Phi, \
-\infty < \rho < \infty .
\end{equation}
Certainly it is possible to prove (\ref{10}) by a simple
substitution $\Phi $ (\ref{11}) into (\ref{8}); for $d=3$
it is the Langer transformation introduced earlier ad hoc.

Taking into account eigenvalues
\begin{equation}
\label{12}
\Delta_{d-1} (\Omega ) Y = - L^2 Y; \quad
L^2= l(l+d-2), \  \ l=0, \ 1, \ 2, \ \dots
\end{equation}
and the term $K(d)$ in (\ref{11}), for
$\Psi = \psi (\rho ) Y (\Omega ) $ we obtain
\begin{equation}
\label{13} \psi'' + W \psi - \lambda^2 \psi = 0, \quad
\lambda = l + \frac{d-2}{2},
\end{equation}
\begin{equation}
\label{14}
W(E, \rho ) = r^2 P^2(r) =
2 e^{2 \rho} \left( E- V(e^{\rho })\right).
\end{equation}

Usual condition $\Phi (r=0) < \infty $ leads to $\psi (\rho ) \to 0$
if $\rho \to - \infty $. Exact spectra of (\ref{8}) and (\ref{10}),
(\ref{13}) must be identical as $\Phi $ and $\Psi $ are only
distinguished by a positive factor $e^{2 \rho } $.

We have seen that the usual "automatic" replacement $l(l+1) \to
\lambda^2 $ actually means that we work in a new special
conformal curved space; its curvature is $K e^{- 2 \rho} $ with
$K$ from (\ref{11}). Instead of the topology $R^d = S^{d-1} \times
(0,\infty ) $ we get $S^{d-1} \times (-\infty , \infty ) $ in the
conformal space.

Coordinates $(\rho , \Omega ) $ are similar to the Cartesian ones in
the maximum possible measure: $\rho $ is a harmonic
coordinate, there is a field of the parallel vectors and
all sections $\rho = const  $  are identical ones \cite{TMP03}. That is why the
leading WKB approximation in the conformal space gives the best
possible result (while exact spectra must be identical in two
metrics $ds^2 $ and $d\tilde{s}^2 $).

It is known \cite{Berezin} that the number of eigenstates with the same
value of $l$ is
\begin{equation}
\label{15}
D(l) = \frac{\Pi_l - \Pi_{l-2}}{(d-1)!};
\quad
\Pi_l = (l+d-1)(l+d-2)\dots (l+1)
\end{equation}
and $\Pi_l=0$ if  $l < 0$.    Substituting in (\ref{15})
$\lambda $ (\ref{3}) we get the leading term $\tilde{D}$:
\begin{equation}
\label{16}
D(\lambda ) = \tilde{D} (\lambda ) + r (\lambda, d);
\quad \tilde{D} = \frac{2 \lambda^{d-2}}{(d-2)!}, \ d \ge 2,
\end{equation}
where $r (\lambda , d) = 0 $ if $d \le 4$ and $r$  is  of  order
$\lambda^{d-4}$ if $d > 4$.

The leading term of $D$ as a function of $l$ is simply $\tilde{D}
(l)$, but $r(l,d)$ is of order $l^{d-1}$. For example, if $d=4$ we
have $D(\lambda ) = \tilde{D} (\lambda ) = \lambda^2 = l^2+ 2l +1
= \tilde{D}(l)+ 2 l +1$. Thus in $D(\lambda ) $ there is no term
of order $\lambda^{d-3}$ unlike $D(l)$, so that namely $\lambda$
is the most suitable variable with smallest distinction
between $D$ and the leading term $\tilde{D}$.

The WKB quantization condition for (\ref{13}) is
\begin{equation}
\label{17}
I (E, \lambda )\equiv \frac{1}{\pi} \int \sqrt{W(E,\rho) - \lambda^2}
d\rho = n_r+ \frac{1}{2} \equiv \nu .
\end{equation}
Obviously (\ref{17}) is identical to (\ref{4}) when we return to previous
variables $r, \ P^2$ (\ref{8}).

\section[3]{\label{sec:section3}
Linear in $\lambda$ approximation of the WKB integral}

As we have already seen, $I (E, \lambda )$ is linear in $\lambda $
for the oscillator and Coulomb potentials. In general case we can write
\begin{equation}
\label{18}
N_1 (E)  \equiv I (E, 0 )=  I (E,\lambda) + \phi \lambda + q (E, \lambda ),
\end{equation}
where $q$ denotes all non-linear corrections. For above cases (\ref{1}),
(\ref{2}) $q \equiv 0$ and $\phi = 1 $ and $\phi =1/2 $ respectively.

For determining $\phi $ we use hereafter the known estimates $N_d$
for the total number of bound states in $d$ dimensional problem
with the energies not exceeding $E$ \cite{Simon}; being expressed
in our variables $W, \; \rho $ they look as
\begin{eqnarray}
\label{19}
N_d & = &
\frac{B\left(\frac{3}{2} , \frac{d-1}{2} \right)}{\pi (d-2)!}M_d
(E); \nonumber \\
M_d (E) &  = & \int W^{d/2} (E, \rho ) d \rho, \quad d \ge 2,
\end{eqnarray}
where $B(y, z)$ is the beta function. Let's show how $N_d$
(\ref{19}) may be obtained using (\ref{17}) and (\ref{16}). At a
fixed $\lambda $ (\ref{5}) the equation (\ref{17}) determines
energies of the bound states. Evidently the maximum value of $\nu
$ corresponds to the maximum energy, so that $I (E, \lambda ) $ is
equal to the total number of bound states with a given $\lambda $
and values of energies not exceeding $E$ (as usual in WKB we
neglect the  difference between $n_r$ and $\nu $ in this  case).
Now we have to take into account the degeneration of states
(\ref{16}). Using the universal form of the first term $\tilde{D}
$, we multiply $I(E,\lambda )$ (\ref{17}) by $\tilde{D}$ and
integrate with respect to  $\lambda  $ over  all possible domain
$0 \le \lambda \le A$,
\[
A^2 (E) = \max\limits_{\rho } W(E, \rho ).
\]
We suppose that $W(E,\rho)$ has sole maximum at any $E$. In
intermediate calculations we treat $\lambda  $ as a continuous
variable. Changing the order of integration we  really obtain
\begin{equation}
\label{20}
\int_0^{A (E)} \tilde{D} (\lambda ) I(E , \lambda ) d\lambda =N_d (E)
\end{equation}
with $N_d$ from (\ref{19}). Thus namely the leading term $\tilde{D} $
but not $D$ must be used in WKB methods.

Now we return to (\ref{18}) and intend to choose value of $\phi $
so that $q (E , \lambda )$ averaged over all bound states becomes zero.
Multiplying both sides of (\ref{18}) by $\tilde{D}$ and integrating with
respect to $\lambda $ we obtain
\begin{equation}
\label{21}
\frac{N_d}{A^d} = \frac{2 N_1}{A(d-1)!} - \frac{2 \phi }{d (d-2)!},
\end{equation}
 Obviously $N_d $ is proportional
to $A^d$ so that the value of $\phi $ does not depend on $A$. It is
also invariant under the transformation $r \to ar $, i.e. $\rho \to
\rho + \ln a , \ a > 0$.

In order to simplify (\ref{21}) we introduce
relative parameters $\chi_d$ comparing $N_d $ for a given potential
with $N_d^{(c)}$ for the above reference case:
the Coulomb potential with the same value of $A$ as in $N_d$.
Calculating $N^{(c)}_d$ we get
\begin{equation}
\label{22}
\chi_d = \frac{N_d}{N_d^{(c)}} =
\frac{d! N_d}{2 A^d}, \quad d \ge 2; \qquad \chi_1 = \frac{N_1}{A} .
\end{equation}
We have used here some identities between the beta functions and their
products. Substituting (\ref{22}) in (\ref{21}) and using (\ref{19})
we obtain finally
\begin{equation}
\label{23}
\phi = \chi_1 + \frac{\chi_1 - \chi_d}{d-1};
\quad
\chi_d =\frac{M_d}{A^d m_d},
\quad m_d = B\left(d/2, 1/2 \right).
\end{equation}
In general case we have $\phi = \phi (d,E)$. For our reference
cases (\ref{1}), (\ref{2}) $\phi = \chi_d = const $ does not
depend on $d$ as well  as on $E$ and coincides with the discussed
above $\phi = 1, \ 1/2 $. Thus we have received the desired
effective quantum number $T$ (\ref{7}) with $\phi $ of a
(\ref{23}). All the parameters $\chi_d$, $\phi$ are some
functionals of a given potential:  $\phi = \phi [V]$ and so on as
well as some functions of $E$ and $d$. Hereafter we indicate only
actual in each case arguments.

The quantization condition (\ref{17}) with respect to (\ref{18}), (\ref{21})
takes the following form:
\begin{equation}
\label{24}
A (E) \chi_1 (E) =\frac{1}{\pi}\int \sqrt{2\left(E-V(r)\right)}dr=
T \left( n_r , l; \phi (E) \right).
\end{equation}

Functions $\chi_d (E) $ may be treated as a special non-linear
transform of a given potential $V(r)$. This transformation is the
most adequate one for our method.

Note that $\phi $ and $\chi_d $ are invariant under the transformation
 $W \to B W$ with $B >0$.

In what follows we treated (\ref{23}) as the basic form for $\phi $.
Nevertheless it is useful to find another approximation for $\phi $
and to compare them in order to evaluate their accuracy. Suppose
we construct a multiplicative expression for $\phi $
\[
\phi_m (E, d) = C N_d^{\alpha } N_1^{\beta }
\]
using only integral estimates $N_d$ and excluding the above
parameter $A$. Since $N_d \propto A^d$ we must put $\alpha d + \beta
= 0$. Substituting $\chi_d$ instead of $N_d$ we get $\phi_m = C'
\left( \chi_1 \chi_d^{-1/d}\right)^{\beta}$. Demanding that
$\phi_m = \chi_d = \phi $ for our reference cases we have finally
\begin{equation}
 \label{25} \phi_m (E,d) = \left(
\frac{\chi_1^d}{\chi_d}\right)^{\frac{1}{d-1}} .
\end{equation}
The same
value of $\phi_m $ we have received earlier using a duality
between certain pairs of power-law potentials \cite{TMP00}. As we
shall see, $\chi_d $ is a smooth monotonic function of $d$ for all
interesting model potentials. It's conveniently to write

\begin{equation}
\label{26}
\chi_d = \chi_{\infty} \exp f_d ,
\end{equation}
introducing an auxiliary function $f_d$; then the ratio of $\phi_m
$ and $\phi $
\begin{equation}
\label{27}
R=\frac{\phi_m}{\phi} = 1 +
\frac{d}{2(d-1)^2}\left(f_1 - f_d \right)^2 + O (f^3) .
\end{equation}
Numerical calculations really give $R > 1$ but very close to $1$,
so that $R-1 < 0.01 $ and in most cases $ < 0.002 $ in a wide
interval of $E$ for a wide set of studied potentials and for all
$d$, see Table \ref{tab:table1}.

This closeness of two values of $\phi $ calculated by quite
different methods confirms the objective character of $\phi $
and of the effective quantum number $T$ itself. We shall return to
this closeness in the section \ref{sec:section7}.

Note that we may treat $d$ in $M_d, \chi_d $ and so on as a
continuous variable in intermediate calculations.

\section[4]{Non-linear transform $\chi_d $ for the
power-law potentials}

In the present section we study power-law potentials
\begin{equation}
\label{28}
V(r) = b r^{\mu}, \quad b \mu >0
\end{equation}
with $-2 < \mu < \infty $. For them  functions $\chi_d $ (\ref{22})
and $\phi $ (\ref{23}) are monotonic and not depending on $E$
\begin{eqnarray}
\label{29}
\chi_d (\mu) & = & \left( \frac{2+\mu}{2} \right)^{\frac{2+\mu}{2}\frac{d}{\mu}}
\frac{2^{d/2}}{\mu^{d/2+1}}
\frac{B(d/\mu ,d/2 +1) }{B(d/2 , 1/2)}
\nonumber \\
 &  & \qquad (\mu > 0), \\
\chi_d (\mu) & = & \left( \frac{2}{2-|\mu |}
\right)^{\frac{2-|\mu|}{2}\frac{d}{|\mu |}}
\frac{2^{d/2}}{|\mu|^{d/2+1}}
\frac{B(d\frac{2-|\mu |}{2|\mu |} ,\frac{d}{2} +1) }{B(d/2 , 1/2)}
 \nonumber \\
 &  & \qquad  (-2 < \mu < 0) \nonumber.
\end{eqnarray}
In the limiting case $d \to \infty $ we obtain from (\ref{28}), (\ref{29})
for all $\mu > - 2 $
\begin{equation}
\label{30}
\chi_{\infty} = \frac{1}{\sqrt{\mu +2}}.
\end{equation}
Naturally (\ref{29}) does not depend on $d$ if $\mu =2,\  \mu = -1 $.
Values of $\chi_d$ and $\phi $ (\ref{23}) are shown in the Table 1.

For the power-law potentials the following expansion$f_d$
(\ref{26}) with $\chi_{\infty} $ (\ref{30}) is valid
\begin{equation}
\label{31} f_d = \sum\limits_{k=1}^{\infty} \frac{b_k}{d^k}, \quad
\chi_d = \chi_{\infty} \left(1 + \frac{b_1}{d} + \dots \right)
 \end{equation}
\begin{eqnarray}
\label{32} b_1 (\mu ) & = & \frac{(\mu+4)^2}{12(\mu +2
)}-\frac{3}{4},  \nonumber \\
b_3(\mu) & = & \frac{1}{360}\left(7+\frac{8\mu^3}{(\mu+2)^2} \right),
\quad b_2 = b_4  \equiv 0 .
\end{eqnarray}
The values of $b_1$ and $b_3$ are equal to
zero for the oscillator and Coulomb potentials and are small in
their vicinity with $|b_3| \ll |b_1|, \ b_1 b_3 < 0$.
Correspondingly the ratio $R$ (\ref{27}) is very close to one. At
any fixed $d$ the value of $\phi ( d, \mu )$ (\ref{23}) with
$\chi_d $ (\ref{29}) decreases monotonically with increasing $\mu
$.

\section[5]{The levels ordering in accordance with $T$ and exact results}

Since the levels ordering is a very important property of
many-particle systems many authors have tried to obtain exact
theorems for several special forms of potentials and only for
$d=3$. On the other hand, the effective quantum number $T$
(\ref{7}) together with the condition (\ref{24}) immediately leads
to the universal for all potentials $V(r)$ ordering
\begin{equation}
\label{33}
sgn \left( E(T) - E(T') \right) = sgn \left( T - T' \right),
\end{equation}
since the left side of the condition (\ref{24}) is a monotonous
function of $E$. In the present section we show that our
expression (\ref{7}), (\ref{23}) for $T$ is exact enough to
reproduce many known or expected results.

It is interesting to compare (\ref{33}) with known results.
The first example connects two differential operators ($d=3$)
\begin{eqnarray}
\label{34} Y_1 [V] & =  & r \frac{d^2}{d r^2}(rV) = r
\frac{dV}{dr}
\left(\kappa(r) +1\right),  \nonumber \\
Y_2 [V] & =  & \frac{d}{dr}\left( \frac{1}{r} \frac{dV}{dr} \right) = r
\frac{1}{r^2}\frac{dV}{dr}
\left(\kappa(r) - 2\right), \\
\kappa (r) & = & 1 + \frac{r\left(d^2 V/dr^2\right)}{dV/dr}
\nonumber
\end{eqnarray}
with two energy differences correspondingly:
\begin{eqnarray}
\label{35}
D_1  & =  & E(n_r + 1, l) - E(n_r, l+1),   \nonumber \\
D_2  & =  & E(n_r, l) - E(n_r - 1 , l+ 2) .
\end{eqnarray}
We have introduced an auxiliary function $\kappa (r)$; for power-law potentials
$\kappa (r)  \equiv  \mu $.

The theorem proved in \cite{Martin87,22} states in our notation
that
\begin{eqnarray} \label{36}
sgn Y_1 & = & sgn D_1,    \nonumber \\
sgn Y_2 & = & sgn D_2
\end{eqnarray}
if $ sgn Y_k = const $ for $0
< r < \infty $.
Suppose our $T$-method is exact. Then, taking
into account that for monotonic attractive
potentials $dV / dr  > 0 $ in (\ref{34}) and using (\ref{33}), we
can write instead of (\ref{36})
 \begin{eqnarray}
 \label{37}
  sgn Y_1 [V] & = & sgn  \left( \kappa (r) + 1 \right)  \nonumber \\
& = &    sgn \left( T(n_r + 1, l) - T(n_r, l+1) \right), \nonumber \\
sgn Y_2 [V] & = & sgn \left( \kappa (r) - 2  \right) \\
 & = & sgn \left(T(n_r, l) - T(n_r - 1 , l + 2) \right). \nonumber
\end{eqnarray}
For
power-law potentials $\phi = \phi [ \mu ] $ does not depend on
$E$,  $\kappa = \mu $ and substituting $T$ (\ref{7}) into
(\ref{37}) we get two equivalent to (\ref{36}) equations
\begin{subequations}
 \begin{eqnarray}
 \label{38}
 sgn \left( \mu  + 1 \right) & = & sgn (1 -  \phi ),
\label{equationa}
\\
 sgn \left( \mu  - 2 \right) & = & sgn (1 - 2 \phi ).
\label{equationb}
\end{eqnarray}
\end{subequations}

The equality (\ref{38}) really takes place for all potentials
(\ref{28}) as $\phi $ (\ref{23}) with $\chi_d $ (\ref{29}) is a
monotonic function of $\mu $, see examples in Table~\ref{tab:table1} , $\phi = 1/2
$ for $\kappa=\mu =2 $ and $\phi =1 $ for $\kappa=\mu = -1 $.

Another exact statement is the following one \cite{Martin80}: for
power-law potentials (\ref{28}) $d^2 E(0,l) /dl^2 > 0$ if $\mu > 2
$. From our condition (\ref{24}) we get immediately
\[
E(n_r , l) =
 C_{\mu } T^{\frac{2\mu}{\mu + 2}} =
C_{\mu } \left( \nu + \phi
\lambda \right)^{\frac{2\mu}{\mu + 2}},
\quad C_{\mu} > 0, \  \mu > 0,
\]
\[
sgn \frac{\partial^2  E}{\partial l^2}  =  sgn ( \mu - 2) ,
\]
so that the abovementioned inequality really fulfills with our $T$
for all $n_r$.

As the third example we study a family of potentials for
quarkonium systems \cite{Martin80}
\begin{eqnarray}
\label{39}
V_q (\alpha ,  \delta , r) & = &
B\left( -\frac{\alpha}{r} + ( 1 - \alpha ) r^{\delta} \right),
\nonumber \\
 & &  0 < \alpha < 1 , \quad \delta > 0, \quad B > 0.
 \end{eqnarray}
 The following ordering  was discussed for $V_q$
\begin{equation}
 \label{40}
   E(0,1)   <   E(1,0) <  E(0,2) .
\end{equation}
For this family  $\partial V / \partial r > 0$, $dk  / dr > 0$,
\begin{eqnarray}
\label{41}
-1 < \kappa (r) = \frac{-1 + Q \mu^2}{1+ Q \mu } & < & \kappa (\infty ) = \delta  ,
\nonumber \\
Q (r)  =  \frac{1-\alpha}{\alpha} r^{\delta + 1}
\end{eqnarray}
for any $r > 0$. Thus the above theorem (\ref{36})
holds and predicts the ordering (\ref{40}), but only for $\delta \le 2 $
(if $\delta > 2,\  sgn (\kappa - 2)$ changes its sign at large $r$).

Now we study how our $T$-method works in this situation.
Since we act within the WKB frame it is natural a "weak" \\
\textit{Conjecture 1.}
The levels ordering satisfies (\ref{37}) if $sgn (\kappa - 2) = const $
only in the classically accessible domain:
\begin{equation}
\label{42}
r < r_t , \qquad V(r_t) = E .
\end{equation}
(we suppose $\partial V / \partial r  > 0$ for our potentials).
This conjecture allows us to have $\kappa (r_t)$ and correspondingly
the previous levels ordering as for $\mu \le 2$ even $\delta > 2$.
It is easy to calculate that for (\ref{39})
$d \kappa (r_t) /d\alpha < 0$, so that we can have $\kappa (r_t) < 2$ for
small values of energy or for middle values of $\alpha $ but
$\kappa (r_t)  > 2 $ and the inverse ordering
$E(1,0) > E(0,2)$ if $B(1-\alpha )$ is great enough.
Rigorous results confirm it \cite{Martin80}.

Besides, we have $d (\kappa_m ) /  d E > 0$.
Note that for $\alpha \ll 1 $ the Coulomb term is negligible
in (\ref{39}) even for the deepest levels as compared with
$B (1 - \alpha) r^{\delta} $.

Moreover, it seems to be reasonable a "strong" \\
\textit{Conjecture 2.}
For any potential with a smooth $d \kappa / dr$
the value of $\phi $ can be well approximated as
$\phi [\mu ]$ for power-law potential (\ref{28}) with
$\mu = \kappa (r_m)$. Here $r_m$ depends both on the value
of the energy and the parameters of a given potential,
and $r_m$ is defined as $W(r_m) = \max$.

Really, this conjecture is exact for $\chi_{\infty}$,
compare (\ref{30}) and (\ref{49}). Each $\chi_d$
and thus $\phi$, as it follows from (\ref{31}) and (\ref{47}),
includes functions $b_k$ depending on some derivatives
$W^{(n)} (r_m) $.
We can express $W^{(n)}(r_m)$, $a_k$ (\ref{47}) and
correspondingly $b_k$ (\ref{31}) by means of $\kappa (r)$ (\ref{34}).
For example, the main terms take the following form
\[
\chi_d = \chi_{\infty} [\mu_m]
\left( 1+ \frac{b_1 [\mu_m ] + b_1^{add}}{d} + \dots \right)
\]
with $\mu_m \equiv \kappa (r_m ) , \  \chi_{\infty} $ accordingly to (\ref{30})
or (\ref{49}),  $ b_1 [\mu ] $ from (\ref{32}), a new function
$b_1^{add} $ depending on $\kappa (r_m )$
in such a way that $b_1^{add} \equiv 0 $ if $\kappa (r) = const$ and
\[
\frac{d \kappa}{d x} = r_m \frac{d \kappa (r_m)}{d r } .
\]
Using (\ref{49} ) we can write for $d=3$ with the same accuracy
\[
\phi  =  \phi (\mu_m ) +\phi^{add},
\]
\[
\phi^{add}  =  \frac{4   b_1^{add} }{3 \sqrt{\mu_m + 2}}
\]
with one of the previous expressions for $\phi $. In the most or even all
real cases we have $\phi^{add} \ll \phi $. So linear in $\kappa_r $ and
neglecting $\kappa_r^2 , \ \kappa_{rr} $ approximation is
\[
b_1^{add} = \frac{16 + \kappa (r_m )} {24 \left( \kappa (r_m) + 2 \right) }
r_m \frac{d \kappa (r_m)}{d r}
\]
Using (\ref{41}) we get
\[
\frac{d \kappa}{ d x} =
r \frac{d \kappa }{d r} \sim \frac{1}{Q(r_m)} \ll 1
\]
at large $Q$. The condition $Q \gg 1$ fulfills at large $E$
 and/or large $B(1-\alpha)$. Thus "non-adiabatic" correction
$b_1^{add} \ll b_1, \ \phi^{add} \ll \phi$.

All the investigated potentials have $\chi_d , \phi $ as a
smooth function both of $E$ and $d$ as well as of
other parameters. It seems to be correct for all
physically well-founded potentials.

Then the coincidence of the two levels
\[
T(0,2) - T(1,0) = 1 - 2\phi
\]
is possible if $\delta > 2$ and $E$ or $B(1 - \alpha )$ is large enough.
So the line $1 - 2 \phi = 0 $ on the plane $\left( \delta  , B ( 1 - \alpha ) \right)$
divides two domains with opposite levels ordering.
Note that $r_m < r_t $; for the above case
\[
r_m \sim \left( \frac{2}{2+\delta} \right)^{\frac{1}{\delta} } r_t < r_t .
\]

\begin{table*}
\caption{\label{tab:table1}
Values of $\chi_d $ and $\phi $ (\ref{23})
for several potentials: (\ref{45}a,b,c),  (\ref{45}d) $V_{TF} (r) $ and (\ref{28}),
as well as $\phi_m (3)$ (\ref{25}) for these potentials.}
\begin{ruledtabular}
\begin{tabular}{lllllllll}
$  $ & $ V(r) $ &  $\chi_{\infty}$ & $\chi_3$ &  $\chi_2 $
& $\chi_1=\phi({\infty} )$ & $\phi(3) $ & $\phi(2) $ & $\phi_m (3)$
\\ \hline
$E=0$    &  (\ref{45}a) & $1.414$ & $1.376$ & $1.359$ &      $1.316$  & $1.286$
 &  $1.273 $  & $1.286$
\\
    $ $  &   (\ref{45}b) & $  2  $ & $  2  $ & $  2  $ &  $  2  $  & $  2  $  &
$  2  $  &  $  2  $
\\
    $ $  &   (\ref{45}c) & $1.826$ & $1.803$ & $1.793$ &  $1.769$  & $1.752$
    & $1.745$  & $1.752$
\\
 $ $  &      (\ref{45}d) & $1.89$  & $1.87$  & $1.84$  &  $1.78$   & $1.74$
 & $1.72$ &    $1.75$
\\   \hline
$E\to - \infty $ & (\ref{45}a,b,c,d) &
$1$  &  $1$ & $1$     & $1$ & $1$ & $1$ & $1$
\\ \hline
 &  $\mu =-1$ & $1$   & $1$   & $1$   & $1$    &
                                 $1$   & $1$   & $1$    \\
 &  $\mu \to 0$ & $0.707$   & $0.688$   & $0.680$   & $0.658$    &
                                 $0.644$   & $0.636$   & $0.643$    \\
  $ V= b r^{\mu}$  &  $\mu =1$ & $0.577$ & $0.568$ & $0.563$ & $0.551$  &
                               $0.543$   & $0.539$ & $0.544$  \\
                   &  $\mu =2$ & $0.5$   & $0.5$   & $0.5$   & $0.5$    &
                                 $0.5$   & $0.5$   & $0.5$    \\
any   $E$          &  $\mu=3$ & $0.447$  & $0.457$ & $0.461$ & $0.469$  &
                                $0.475$  & $0.477$ & $0.476$  \\
           &  $\mu\to \infty$ & $0    $  & $0.212$ & $0.250$ & $0.318$  &
                                $0.371$  & $0.386$ & $0.390$   \\

\end{tabular}
\end{ruledtabular}
\end{table*}

\section[6]{Screened Coulomb potentials}

In the present section we study with the help of our new method
another actual and interesting class of centrally symmetric potentials
\begin{equation}
\label{43}
V(r) = -\frac{Z g(r)}{r}; \quad g(0) = 1, \ g > 0, \ \frac{d g}{d r} <   0 .
\end{equation}
The Thomas-Fermi potential $V_{TF} (r)$ of the selfconsistent field in the
many-electron atoms \cite{Gombas}
also belongs to the type (\ref{43}). All such model atomic
potentials must obey the inequality $ g'' > 0$. It follows from the
Poisson equation
\[
\Delta U = \frac{Y_1 [U]}{r^2} =- 4\pi \rho
\]
for the electrostatic potential $U$, so that $V=-|e|U$ and the electron
charge density $\rho < 0$. Obviously $sgn Y_1 [V] = - sgn Y_1 [U] = -1$,
correspondingly we get $\kappa < -1 $ from (\ref{34}).
On the other hand, for (\ref{43}) we obtain
\begin{equation}
\label{44}
\kappa = -1 + \frac{g'' r^2}{g' r - g} = -1 - \frac{g'' r^2}{|g' r| +|g|}
\end{equation}
with respect to the inequalities (\ref{43}), so that actually $g'' > 0$.

There are also important potentials with
\begin{subequations}
\label{45}
\begin{eqnarray}
g(r) & = & e^{-r}, \label{45a}
\\
g(r) & = & \frac{1}{(1+r)^2},  \label{45b}
\\
g(r) & = & \frac{1}{(1+r)^{2.5}} \label{45c}
\end{eqnarray}
\end{subequations}
besides (\ref{45}d) $V_{TF} (r)$.

For all these potentials the values $\chi_d $ and $\phi(d)$ depend on the
value of the energy $E$. For the deepest levels where only a small
domain $r < r_t \to 0$ is classically accessible in (\ref{24})
($V(r_t) = E$), we have
$\kappa \to -1 $, and $\chi_d \to \phi(d) \to 1$. Of course in our quantum
problem the deepest level has some small $r_t \ne 0 $ as well as $E\ne -\infty $,
so that formally we have a very small distinction from this limiting
values of $\chi_d $, $\phi (d)$.

In the opposite extreme case $E\to - 0$ corresponding values $\chi_d $ and
$\phi (d)$  for (\ref{45a}),  (\ref{45b}), (\ref{45c})
and (\ref{45}d) $V_{TF}(r)$ are shown in the
Table \ref{tab:table1}. These values asymptotically coincide in
the depth of the potential well where is no screening as the
turning point $r_t \to 0$ for these $V(r)$.
In the same Table we have placed $\chi_d $ and $\phi (d)$ for
some power-law potentials (\ref{28}), including $\mu \to \infty $
(the rectangular potential well) as well as multiplicative
$\phi_m (3) $ from (\ref{25}).

Let's demonstrate two ways to use $T$ (\ref{7}) taking as an example
the model  potential (\ref{45}c) which is a very good
approximation of the real selfconsistent atomic potential
\cite{JETP} with $Z$ being the nuclear charge. If we fix $E=0$
then (\ref{24}) indicates the order in which new bound states
appears with increasing $Z$ as well as corresponding values of
$Z$. On the contrary at $Z=const$ we get values and thus the level
succession of all the bound states in a given atom with a fixed $Z$.
It can be easily seen from our table and (\ref{24}) that shallow
levels $(E \approx 0)$ are governed by $ n_r + 1.75 l$  (of course
$d=3$) but deepest levels by $n_r + l$ i.e. are Coulomb-like, with
an intermediate behavior for middle levels. The introduced
$T$-ordering formalizes the well known quantitative picture and in
particular explains the Periodic system of the elements,
see the next section.

Our method is valid and simple not only for potentials
with non-trivial analytic form as (\ref{45c}) but also for
potentials given numerically. So for the Thomas-Fermi atomic
potential we obtain $\phi (3) $ very close to the corresponding
value for (\ref{45c}).

\section[7]{\label{sec:section7}Universal diagram}

\begin{figure}
\includegraphics{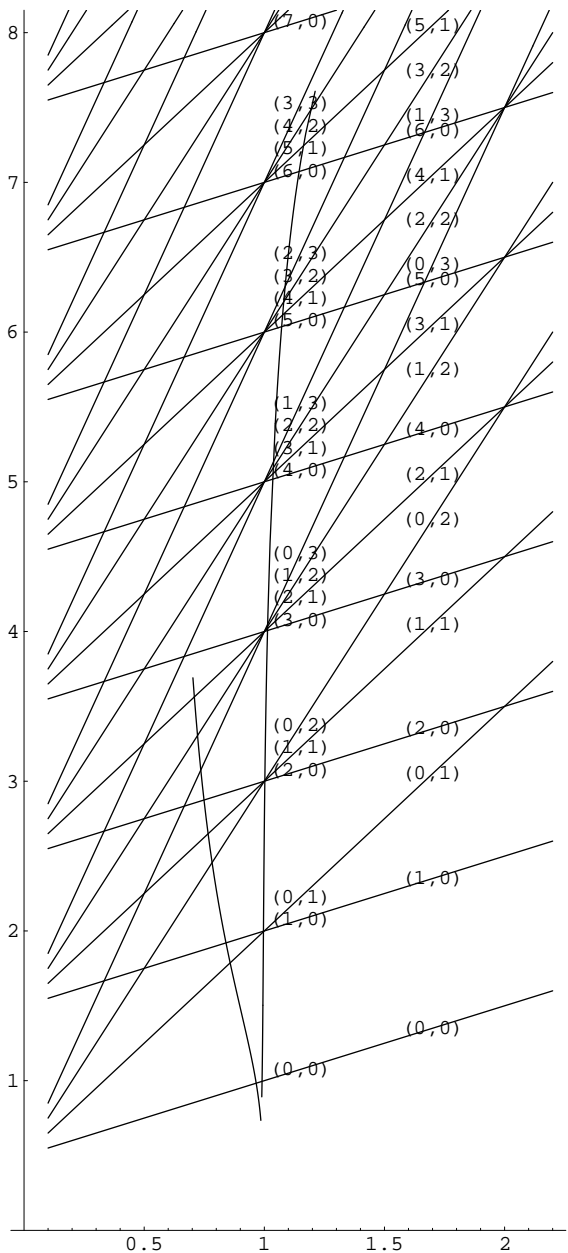}
\caption{\label{fig:diag}
The effective quantum number $T$
(\ref{7}) versus $\phi $ (\ref{23}) for the bound states $(n_r,l)$
with $n_r$ and $l$ being the radial and orbital quantum number
respectively. Lines with $l > 3$ are omitted for the sake of
simplicity.}
\end{figure}

The regular filling of shells in a centrally symmetric system with
an arbitrary dimensionality and nature of the selfconsistent field is
clearly described by the following diagram  Fig. \ref{fig:diag}.
Here each line represents $T(n_r,l,\phi ) $ as the linear function
of $\phi $ at fixed $(n_r , l ) $. Crossing of
two lines marks the values of $\phi $ at which the order
of the level succession changes. Two different types of problems may
be treated with the help of this diagram.

If $E=0$ the value of $\phi$ is invariant under the transformations
$V \to cV, \ r \to c_1 r$, this value does not depend on $Z$ (even if
$g = g \left( r f(Z) \right)$; that's not the case if $E\ne 0$).
With increasing strength of
the potential (i.e. $Z$ in (\ref{46}) ) new shells $(n_r, l)$
with $E=0$ appear in the order of increasing $T(n_r,l)$
at a given value of $\phi (E=0)$, i.e. along the vertical line
$\phi = const $. Each shell can contain $D(d)$
(\ref{16}) states; only this number $D$ depends on $d$.  When
the shell $(n_r , l)$ is filled in, i.e. all states are occupied with
particles,  begins filling of the next
shell $(n_r', l')$ with the value $T' > T$ nearest to $T$.

It is known that the structure of the Periodic system of the
elements corresponds to the definite order of the atomic shells
filling with increasing $Z$ \cite{Ostr2001,JETP}. The actual
shells order is:   $1s,\ 2s, \ 2p, \ 3s, \  3p, \ 4s, \ 3d,  \ 4p,
\ 5s, \ 4d, \ 5p, \ 6s, \ 4f$ i.~e. $(0,0), \ (1,0),  \  (0,1)$
etc. It takes place if and only if
\begin{equation}
\label{46}
 5/3   < \phi (3) < 2
\end{equation}
as it follows from Fig.~\ref{fig:diag}; for the first time this range was
found out in another equivalent form of the atomic potential
asymptotics \cite{JETP, TMP99}.
The most probable value $\phi \approx 1.75$ corresponds to the
Thomas-Fermi potential as well as
to (\ref{45c}) and satisfies (\ref{46}).

The usually supposed shells ordering in metallic clusters
\cite{Ostr97} means in our notation that $ \phi = 1/3 $. The close
value $ \phi (3) \approx 0.37  $ for all levels we obtain for the
rectangular potential well (\ref{28}) with $\mu \to \infty $.

We can also add that the $T$-ordering or immediately Fig.~\ref{fig:diag} exactly
reproduces the levels ordering calculated for (\ref{28})
with $d=3$ and several values of
$\mu $, e.~g. for $\mu = 0.1 $, $\phi \approx 0.63 $ \cite{21}
and $\mu \to \infty $, $\phi \approx 0.37$.

Another situation arises if we treat $\phi (E)$ and the left side of
the condition (\ref{24}) $A(E) \chi_1 (E)$ at a fixed $E$ as a point
$\left( \phi (E),\;  A(E)\chi_1 (E) \right) $
on the plane of the diagram. Changing $E$ we receive a curve, the form
of which depends on a given potential $V(r)$.

At an arbitrary point we can't find integer $(n_r, l)$ satisfying
(\ref{24}). Such integer $(n_r,l)$ only exist for these
points where one of the lines
$T(n_r , l, \phi )$ crosses the curve. Thus these distinguished points
indicate the actual bound states and indirectly their energies. The
curve as the whole shows clearly how the levels ordering changes
with changing $E$ or $T$ (remind that $dT /dE > 0$).

As an example we indicate on the diagram such curves for the Yukawa
potential $V(r) = - 50 e^{- r} /r$ (the right curve) and the quarkonium
potential $V(r)=  3 \left( -1/r + r \right)$  (the left curve).

\section[8]{Asymptotical behavior of $\chi_p $ and $\phi $ and a
non-linear quantization condition}

As we have demonstrated in previous sections, our method with the linear
in $\lambda $ approximation for $T$ is sufficiently exact. Meanwhile, it is
possible to take into account non-linear corrections. In the present section
we introduce a non-linear form for our effective quantum number $T$
and demonstrate its new possibilities. Let's use the fact that
the integral (\ref{19}) has the original form for the asymptotic Laplace
expansion:
\begin{eqnarray}
\label{47}
M_d & = & \int dx W^{d/2} (x)   \nonumber \\
  & \approx &  A^{d+1} \sqrt{\frac{4\pi }{|W^{(2)}| d}}
\left( 1 + \frac{a_1}{d}  + \frac{a_2}{d^2} + \dots \right) ,
\end{eqnarray}
where $a_k$ are known functions of
the derivatives $W^{(n)} $ taken at the maximum point $r=r_m$,
$A^2 = W (r_m)$.
Expanding $m_d$ in the denominator of (\ref{23}) similarly to (\ref{47})
we obtain an asymptotic expansion of $\chi_d $ (\ref{26}), (\ref{31}), where
\begin{equation}
\label{48}
\chi_{\infty} = A \sqrt{\frac{2}{ |W^{(2)}| }  }.
\end{equation}
A special case of such expansion is (\ref{32}). Of course we can also
calculate $\chi_{\infty} $ as $\lim \chi_d$ for $d \to \infty $ (even for
nonanalytic potentials, when (\ref{48}) may be incorrect).
It is easy to prove numerically that all interesting model
potentials including those represented in the Table~\ref{tab:table1} have
$b_1 \ll 1 $, $|b_2 | \ll |b_1 | $ and so on.
E.~g. for power-law potentials (\ref{28}) we have accordingly to
(\ref{32}) $b_1 (0) = - 0.08, \ b_1(4) = 0.139$ and only in an irreal case
$\mu \approx 15$ we get $b_1(\mu ) \approx 1$.
Thus the asymptotic regime is already reached for $d \ge 1 $
(actually for  $d \ge 0.5 $). That's why $\chi_d $ are smooth monotonic
functions for $d \ge 1$.

Using $\kappa (r) $ (\ref{34}) we can get an interesting expression
equivalent to (\ref{48}):
\begin{equation}
\label{49}
\chi_{\infty} = \sqrt{\frac{2}{\kappa (r_m)+ 2}}.
\end{equation}
For power-law potentials $\kappa \equiv \mu $ and we return to (\ref{30}).

Neglecting in $f_d $ (\ref{26}), (\ref{31})
all terms with $d \ge 2 $ we obtain a simple approximation
\begin{equation}
\label{50}
\chi_d \approx \chi_d^{(as)} = \chi_{\infty} + \frac{\chi_1 - \chi_{\infty}}{d}.
\end{equation}
Substituting (\ref{50}) in (\ref{23}), we also have
\begin{equation}
\label{51}
\phi(d) \approx \phi^{(as)} (d) = \chi_1 + \frac{\chi_1-\chi_{\infty} }  {d}.
\end{equation}

At last comparing (\ref{50}) and (\ref{51}) we obtain another approximate
expression
\begin{equation}
\label{52}
\phi (d) \approx \chi_D^{(as)}, \qquad D=\frac{d}{d+1}.
\end{equation}
We have calculated ratios $s= \phi^{(as)} (d)/\phi(d) $ and
$\chi_D^{(as)} / \phi (d) $ where $\phi (d) $ is the basic form (\ref{23}) for
various potentials and for a wide interval of $E$. It turns out that even for
$d=3$ both $s$ and $w$ are close to unity: $|s-1|, \ |w-1| \le 0.02$ and in
most cases $\le 0.01$.

As we have already said, the closest approximation to the basic additive
form $\phi $ is the multiplicative form $\phi_m$ (\ref{25}): for their ratio
$R= \phi / \phi_m $ we have $R - 1 < 0.01$ and in most cases $< 0.002$,
see Table~\ref{tab:table1}.

Each of these approximations may be preferable in
some suitable situation.

A simple universal form $\chi_d^{(as)} $ (\ref{50}) allows us to get
an universal non-linear approximation
\begin{equation}
\label{53}
I(\lambda ) = I( 0 ) -  F (\lambda )
\end{equation}
for $I (E, \lambda ) $ (\ref{17}). Acting as
in Sec.~\ref{sec:section3} we obtain
instead of (\ref{23})
\begin{equation}
\label{54}
\frac{d}{A^d} \int_0^A F (\lambda ) \lambda^{d-2} d \lambda =
\frac{d \chi_1 - \chi_d }{d-1} .
\end{equation}
If we assume $F = \phi \lambda $ in (\ref{54}), we return to
$\phi $ (\ref{23}). But this time we use $\chi_d $ (\ref{50})
and thus come to a simple Melline transform for
$F (\lambda ) \lambda^{-1} $:
\begin{eqnarray}
\label{55}
& & \frac{1}{A^d} \int_0^A  F   (\lambda ) \lambda^{d-2} d \lambda  =
\frac{\chi_1}{d} + \frac{\chi_1 - \chi_{\infty}}{d^2}, \nonumber \\
& &  F (\lambda ) = \chi_1 \lambda +
\left( \chi_1 - \chi_{\infty} \right) \lambda \ln (\lambda /A) .
\end{eqnarray}
Correspondingly we obtain a non-linear in  $\lambda $ quantization
condition
\begin{equation}
\label{56}
A(E) \chi_1 (E) = T_{non} (n_r , l),
\qquad
T_{non} (n_r , l) = \nu + F (\lambda ).
\end{equation}
Both (\ref{55}) and (\ref{56}) become incorrect if $\lambda /A \ll 1 $,
but actually if $d \ge 3 $ we have a finite $\lambda /A$ even for
$l =0$, see (\ref{13}); in real systems usually
$A \approx \lambda_{max} \le 5$.

Let's reproduce a  delicate distinction
obtained for the power-law potentials
(\ref{28}) by means of the $\hbar$-expansion of the Regge trajectories
\cite{23}. Using our notation we can rewrite the final result \cite{23}
in a simple form for $\mu > -1$
\begin{eqnarray}
\label{57}
& & sgn \left( E^{\sigma} (0,l+1)   -   2 E^{\sigma } (0,l)
+  E^{\sigma } (0,l-1) \right)  = \nonumber \\
 & & = sgn (2 - \mu ),\quad \sigma = \frac{2 \mu}{\mu +1} .
\end{eqnarray}
Since $E \propto T^{\frac{1}{\sigma}} $ for (\ref{28}) the left side of
(\ref{57}) is linear in $T$; by substituting $T_{non}$ (\ref{56}) in
(\ref{57}), we obtain
\begin{eqnarray}
\label{58}
& & sgn \left( \chi_{\infty} - \chi_1 \right)  \cdot  sgn \Lambda
 =  sgn \left( \chi_{\infty} - \chi_1 \right)  \nonumber \\
 &  &  \quad  = sgn (2 - \mu  ), \\
& & \Lambda =
(\lambda + 1 ) \ln (\lambda + 1) +
(\lambda - 1 ) \ln (\lambda - 1) -
2\lambda \ln \lambda  \nonumber .
 \end{eqnarray}
We have used here the fact that $d^2 (x \ln  x) /dx^2 > 0$,
so that $\Lambda > 0, \ sgn \Lambda =1$. But the last equality
(\ref{58}) really fulfills for our $\chi_d $, see e.~g.
Table~\ref{tab:table1}.
Should be stressed that in any linear approximation the left side
of Eq.~(\ref{57}) is equal to zero, and not to one as for $\Lambda$,
so that only non-linear approximation confirms the strong result \cite{23}.

\section[9]{Conclusion}

Thus we have constructed and calculated the effective quantum number $T$
(\ref{7}) which determines with  very high accuracy appearing and ordering
of the bound states in any centrally symmetric potential. It should be stressed
that different potentials may have very close or coinciding values of $\phi $
and $T$ and hence identical levels ordering (see e.~g. cases b) and c) in the
Table~\ref{tab:table1}).

Some partial success of  using the Thomas-Fermi potential to explain the Periodic
system does not mean that this potential is the genuine or the best one. The
point is that its $\phi $ is situated not too close to the limiting points of
 the interval (\ref{46}) so that various
corrections not taken into account in the Thomas-Fermi approach
 though change $\phi $ but remain it
inside this interval. Hence the levels ordering is really the same as for the
Thomas-Fermi potential.

The effective number $T$ actually replaces the principal
quantum number $n = n_r + l +1$ for all potentials besides the Coulomb one
with $d=3$.

Using $T$ we immediately reproduce many results received for the levels ordering by
means of special theorems and numerical calculations. Moreover,
 the quantization condition (\ref{24}) with $T$ can also be used for
determining spectra. The accuracy of the energy values calculated
for $V(r) = r$ is $0.3 \div 0.5 $\% and even in the worst case of the non-analytic
potential well ($\mu \to \infty$) errors do not exceed $3 \div 5$\% \cite{Basel}.
For two potentials with the same value of $\phi $ we usually obtain different
$A$  and $\chi_1$, so that their energies of the bound states do not coincide
unlike the levels ordering.

A wide variety of  the physically interesting potentials, quite different as
transforms of $r$,  have an universal asymptotic
behavior if we use special adequate transforms $\chi_d $ and $\phi $.

\end{document}